\documentclass[onecolumn,showpacs,preprintnumbers,amsmath,amssymb]{revtex4}
\usepackage{graphicx}
\def \be{\begin{equation}}
\def \ee{\end{equation}}
\def \H{{\cal H}}
\def \a{\alpha}
\def \b{\beta}
\def \f+{\vec{f_{+}}}
\def \fx{\vec{f_{\times}}}

\def \X{X_{\rm switch}}

\begin{document}


\title{Optimising the directional sensitivity of LISA}
\author{K.~Rajesh~Nayak$^1$, S.~V.~Dhurandhar$^1$, A.~Pai$^2$ and J-Y~Vinet$^2$}
\affiliation{$^1$IUCAA, Ganeshkhind, Pune - 411 007, India. \\
$^2$CNRS, Observatoire de la C\^ote d'Azur\\
UMR6162--ILGA (Interf\'erom\'etrie Laser pour la Gravitation et l'Astrophysique)\\
BP 4229 F-06304 Nice Cedex 4 France}
\date{\today}
\begin{abstract}
It was shown in a previous work that the data combinations canceling laser
frequency noise constitute 
a module - the module of syzygies. The cancellation of laser frequency noise is crucial for obtaining the requisite 
sensitivity for LISA. In this work we show how the sensitivity of LISA can be optimised for a 
monochromatic source - a compact binary - whose direction is known, by using appropriate data 
combinations in the module. A stationary source in the barycentric frame appears to move in the LISA frame 
and our strategy consists of {\it coherently tracking} the source by appropriately {\it switching} the data 
combinations so that they remain optimal at all times. Assuming that the polarisation of the source is not known, 
we average the signal over the polarisations. We find that the best statistic is the `network' statistic, in which case LISA 
can be construed of as two independent detectors. We compare our results with the Michelson combination, which has been 
used for obtaining the standard sensitivity curve for LISA, and with the observable obtained by optimally switching the 
three Michelson combinations. We find that for sources lying in the ecliptic plane the improvement in SNR increases from 
34$\%$ at low frequencies to nearly 90$\%$ at around 20 mHz. Finally we present the signal-to-noise ratios for some 
known binaries in our galaxy. We also show that, if at low frequencies SNRs of both polarisations can be measured, the 
inclination angle of the plane of the orbit of the binary can be estimated. 
  
\end{abstract}


\maketitle

\section{Introduction}
 The goal of the LISA mission \cite{LISArep} is to detect and analyze low frequency 
gravitational signals essentially in the range from 0.1 mHz to 0.1 Hz. In this
frequency range, sources of gravitational waves are mainly the wide 
population of galactic compact binaries, and interactions between black holes with a range of complexity. 
The study of the emission of GW from  
known binaries could be extremely useful for firstly,  direct determination of distances, 
and secondly, possible small general relativistic effects, if the signal to noise
ratio (SNR) is large enough. For this reason, we focus 
on optimising the sensitivity of LISA for a given astrophysical source 
with known direction. The sensitivity of LISA can be improved by solving
technological problems, but it can also be improved by employing certain optimal data analysis
strategies. In this paper, we show how algebraic methods previously developed in  
\cite{SNV02}, henceforth referred to as paper I, can be used to design optimal strategies for combining
data with appropriate time-delays. We show in this paper that it is possible to maintain 
the optimality during the year by continuously updating the parameters of
the combination or simply by switching to optimal data combinations as the source appears 
to move in the LISA frame, as LISA moves in its complex orbit around the sun. 
The problem of optimisation of SNR, in this context, has been addressed before in \cite{PTL02} and 
\cite{SNV03}. In \cite{PTL02} optimisation has been carried out before averaging over the directions and 
polarisations, while in \cite{SNV03}, henceforth referred to as paper II, the averaging is done first and 
then the optimisation. In this paper, the averaging over the polarisations is performed first and then the 
SNR is optimised for the average signal for a given direction over the relevant data combinations - 
those data combinations canceling laser frequency noise. Thus in this optimisation, the direction of source 
is assumed to be known, but not its polarisation. This would be the case for several binaries in our galaxy.
We analyse two observables in this context: (i) optimal data combination in the module which yields the maximum 
SNR for a given direction and (ii) a `network' 
observable which is obtained by squaring and adding the SNRs of two independent(orthogonal) data combinations one 
of them being the optimal combination mentioned in (i) and another orthogonal to it. The network observable  
in general yields higher SNR. We analyse how these SNRs depend on direction 
of the source. For an integration time of one year, we compare our results by plotting sensitivity curves 
with those obtained from, 
(a)  the Michelson combination $X$ \cite{schilling}, which has been used for obtaining the standard sensitivity  curve for LISA, and 
(b)  the observable $\X$ obtained by switching the three Michelson combinations optimally. 
We find that for sources lying in the ecliptic plane, the network sensitivity at low frequencies is about 34\% 
more than the optimally switched Michelson combinations which rises to nearly 90\% at 20 mHz. 
Finally we compute the optimal SNRs for six known binary systems in our galaxy, whose SNRs are significantly high, 
for an integration time of one year. We show that if at low frequencies the two SNRs of the orthogonal data combinations 
can be measured, then it is possible to estimate the inclination angle of the binary's orbit. 

The paper is organised as follows: In section \ref{module} we briefly describe the formalism involving 
commutative algebra developed in paper I. We also present a convenient set of generators for the data combinations 
found in paper II, which consisted of the eigenvectors of the 
noise-covariance matrix, and in which the computations are simplified. In section \ref{signal} we discuss the 
gravitational wave (GW) signal from non-spinning compact binaries and average the signal over the polarisations. 
In section \ref{SNR} we obtain the optimum SNR by the method of Lagrange multipliers, where we must solve an  
eigenvalue problem. In section \ref{track} we describe the transformations from the LISA frame to the barycentric frame 
and vice-versa. In section \ref{sens} we compute the optimal SNRs and the sensitivity curves and compare them with 
the standard curve for LISA obtained with the Michelson data combination $X$ and the optimally switched Michelson combination $\X$. 
In this section we also estimate the SNRs for a few known binaries in our galaxy and describe a method to estimate the 
angle of inclination of the binary's orbit, if the two SNRs of the orthogonal combinations can be measured.    

\section{The module of syzygies and its generators}
\label{module}

\begin{figure}

\caption{The schematic LISA configuration}

\includegraphics[width=.5\textwidth]{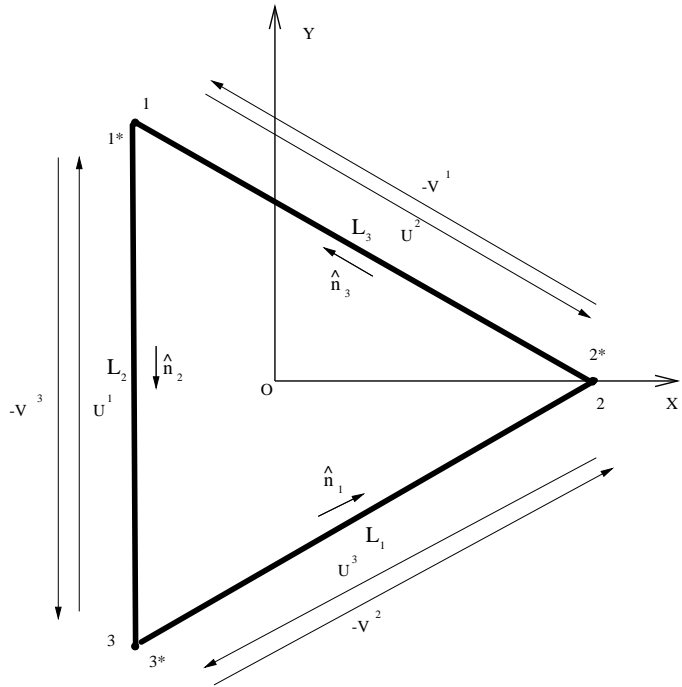}
\label{lisagm}
\end{figure}

In this section, we briefly review the results from papers I and II which are needed for the analysis that 
follows. In LISA the six elementary data streams are labeled as $U^i$ and $V^i, i = 1, 2, 3$; if the space-crafts 
are labeled clockwise, the $U^i$ beams also travel clockwise, while the $V^i$ beams travel counter-clockwise.
The beam $U^1$ travels from space-craft 3 to space-craft 1 along the arm of length $L_2$ in the 
direction $- \hat{\bf n}_2$, while $-V^1$ represents the beam traveling from space-craft 2 to space-craft 1 
along the arm of length $L_3$ in the direction of $\hat{\bf n}_3$. The remaining 4 beams are described by cyclically 
permuting the indices. These beams contain the laser frequency noise, other noises such as optical path, acceleration etc. 
and also the GW signal. For performing the analysis, we choose a frame $(x_L, y_L, z_L)$ tied to the LISA constellation.  
We choose the centroid of the LISA triangle as the origin, the LISA triangle to lie in 
the $(x_L, y_L)$ plane and space-craft 2 to lie on the $x_L$-axis. We will henceforth drop the suffix `$L$' from 
the LISA frame quantities and only introduce it when other frames are being considered in the discussion.
A schematic diagram of LISA is shown in Fig.{\ref{lisagm}}.

In paper I, we showed that all data combinations canceling laser frequency noise form a module over the 
ring of polynomials in three indeterminates, these being the three-time delay operators $E_i, i = 1, 2, 3$ of the 
light travel time along the three arms with lengths $L_i$ respectively. Thus for any data stream $X(t): 
E_i X(t) = X(t - L_i)$. For the LISA specifications $L_i \sim 16.7$ sec. corresponding to an armlength of 5 million 
km.  A general data combination is a linear combination of the elementary data streams $U^i, V^i$:
\begin{equation}
    X(t) = \sum_{i=1}^{3}   p_i V^i (t) + q_i U^i (t) ,
\end{equation}
where $p_i$ and $q_i$ are polynomials in the time-delay operators $E_i, i = 1, 2, 3$. Thus any data combination 
can be expressed as a six-tuple polynomial `vector' $(p_i, q_i)$. For cancellation of laser frequency noise 
only the polynomial vectors satisfying this constraint are allowed and they form the module of syzygies mentioned 
above. An important advantage of this formalism is that, given the generators of the module, any other data combination 
canceling the laser frequency noise is simply a linear combination of the generators with the coefficients being 
polynomials in the ring.

Several sets of generators have been listed in paper I. Based on the physical application 
as also on convenience, we may choose one set over another. 
For the monochromatic sources, as argued in paper II, only three generators are sufficient to generate the relevant 
data combinations. In general 4 generators are necessary to generate the module but if the source is monochromatic, 
except at certain frequencies which are solutions of $e^{i (L_1 + L_2 + L_3) \Omega} = 1$, the fourth generator can 
be effectively eliminated for the purposes of maximising the SNR. Since this maximisation is possible arbitrarily 
close to the singular frequencies the singularities do not seem to be important. 

We found in paper II that a convenient set of generators for our purpose is the one that diagonalises the 
noise covariance matrix. These generators are the eigenvectors of the noise covariance matrix. The noise covariance 
matrix ${\cal N}^{(I)}_{(J)}, I, J = 1, 2, 3$ is defined as the outer product  
${\cal N}^{(I)}_{(J)} = N^{(I)} N_{(J)}^{*}$ where $N^{(I)}$ is a 12 dimensional complex noise vector of the generator 
$X^{(I)} = \left(p^{(I)}_i, q^{(I)}_i\right)$ (The $X^{(I)}$ could be any set of generators not necessarily the eigenvectors of 
the noise covariance matrix ${\cal N}^{(I)}_{(J)}$ ). The noise vector is given by:
\begin{eqnarray}
N^{(I)} &=& \left(\sqrt{S^{pf}} (2 p^{(I)}_i + r^{(I)}_i),\sqrt{S^{pf}} (2 q^{(I)}_i + r^{(I)}_i), 
 \sqrt{S^{opt}} p_i^{(I)}, \sqrt{S^{opt}} q_i^{(I)}\right),
\end{eqnarray}
where the $r_i$ polynomials are defined through the equations $r_1 = - (p_1 + E_3 q_2) = - (q_1 + E_2 p_3)$ plus 
cyclic permutations for $r_2$ and $r_3$. The $S^{pf} = 2.5 \times
10^{-48} (f/1 {\rm Hz})^{-2} {\rm Hz}^{-1}$ and $S^{opt} = 1.8 \times 10^{-37}
(f/1 {\rm Hz})^2 {\rm Hz}^{-1}$ are the one-sided power spectral 
densities (psd) of the proof-mass noise and the optical-path noise respectively \cite{LISArep,ETA00}.
       
For the purposes of this paper when computing the SNR for any given data combination, the differences in armlengths 
are extremely small compared to the wavelength of the GW we are interested in detecting. Thus we may take the 
armlengths to be equal, i.e., $E_1 = E_2 = E_3 = E = e^{i \Omega L}$. For the purposes of this analysis it is sufficient 
to consider LISA as an equilateral triangle; we ignore the deviations arising from this assumption. This simplifies the 
expressions for the noise and further also for the signal. 
The eigenvectors now contain overall common factors which are polynomials in 
$E$. These common factors make no difference to the computation of SNR as they cancel out from the numerator and 
denominator which comprise of the signal and the noise respectively. The unnormalised eigenvectors (with common factors 
canceled) of the noise covariance matrix is the set $\{Y^{(I)}\}$ which we list below:

\begin{eqnarray}
Y^{(1)} & = & \left( 1 - E,\, 1 + 2E,\, -2 - E,\, 1 + 2E,\, 1 - E,\, -2 - E \right)\,,  \nonumber \\
Y^{(2)} & = & \left( -E -1,\, 1,\, E,\, -1,\, 1 + E,\, -E \right)\,,\\
Y^{(3)} & = & \left(1,\,1,\,1,\,1,\,1,\,1\right). \nonumber
\label{basis}
\end{eqnarray}
In paper II, the $Y^{(I)}$ have been listed with the uncanceled common factors.

We adopt the following terminology: we refer to a single element of the module
as a data {\it combination}; while a function of the elements of the 
module, such as taking the maximum over several data combinations in the
module or squaring and adding data combinations belonging to the module,  
is called  as an  {\it observable}.  The important point to note is
that the laser frequency noise is also suppressed for the observable although
it may not be an element of the module.

\section{The GW signal from binaries}
\label{signal}

\subsection{The waveform}

Since binaries will be important sources for LISA the analysis of such sources is relevant. One such class 
is of massive or super massive binaries whose individual masses could range from $10^3 M_{\odot}$ to $10^8 ~ 
M_{\odot}$ and which could be around a Gpc away. Another class of interest are known binaries within our 
own galaxy whose individual masses are of the order of a solar mass but are just at a distance of a few kpc or less.
Here we will assume the direction of the source to be known, which is justified for known binaries in our galaxy; 
but even for the former case of distant binaries, it amounts to `looking' in a specific direction.

The spacetime metric perturbation for a gravitational wave propagating
in a $\hat{w}$ direction can be written as,
\begin{eqnarray}
h_{ij}(t,\vec{r})&=& h_{+}(t-\hat{w}\cdot\vec{r})\left(\theta_{i}\theta_{j}-\phi_{i}\phi_{j}\right) \nonumber \\
        && +h_{\times}(t-\hat{w}\cdot\vec{r})\left(\theta_{i}\phi_{j}+\theta_{j}\phi_{i}\right)\,,
\end{eqnarray}
where the direction to the source is described by direction angles $\left(\,\theta,\,\phi\,\right)$, 
the unit vector pointing to the source is $\hat{w} = (\sin \theta \cos \phi, \sin \theta \sin \phi, \cos \theta)$ 
and $\hat{\theta} $ and $\hat{\phi}$ are unit vectors in the direction of the coordinates $\theta$ and $\phi$, 
such that $\{\hat{w}, \hat{\theta}, \hat{\phi}\}$ form a right-handed triad.

For a binary that is not chirping, and hence monochromatic with frequency $\Omega$, the two polarisation 
amplitudes $h_{+} (t)$ and $h_{\times} (t)$ \cite{SNV02} are given by,  
\begin{eqnarray}
h_{+}(t) & = & H\left[\frac{1+\cos^{2}\epsilon}{2}\cos2\psi\,\cos\Omega t+\cos\epsilon\,\sin2\psi\,\sin\Omega t\right]\,, \nonumber \\
h_{\times}(t) & = & H\left[-\frac{1+\cos^{2}\epsilon}{2}\sin2\psi\,\cos\Omega t+\cos\epsilon\,\cos2\psi\,\sin\Omega t\right]\,,
\end{eqnarray}
where the angles $(\epsilon, \psi)$ describe the orientation of the binary orbit ($\epsilon, \psi$ could be the  
direction angles of the orbital angular momentum vector) and $H$ the overall amplitude which depends on the 
masses, the distance and the frequency $f$ as given below:
\begin{equation}
H=1.188\times10^{-22}\cdot\left[\frac{{\cal M}}{1000\, M_{\odot}}\right]^{\frac{5}{3}} \cdot 
\left[\frac{R}{1\,\mathrm{Gpc}}\right]^{-1}\cdot\left[\frac{f}{1\,\mathrm{mHz}}\right]^{2/3}\,, 
\end{equation}
where $R$ is the distance to the source, $f=\Omega/2\pi$ is the GW frequency of the source and 
${\cal M}=\left(\mu^{3}M^{2}\right)^{\frac{1}{5}}$ the chirp-mass, where $\mu$ and $M$ are respectively the 
reduced and the total mass of the binary. For the typical parameters taken above the frequency evolves very 
slowly, so much so that the time for the system to coalesce is more than 20 years. For an observational time 
of the order of an year, the frequency of the binary changes very little so that the source can be practically  
taken to be monochromatic. 
 
Since we deal with essentially monochromatic sources, the Fourier domain is appropriate for further analysis. 
In the Fourier domain we have,

\begin{eqnarray}
h_{+}(\Omega) & = &  H \left[\frac{1+\cos^{2}\epsilon}{2}\cos2\psi-i\,\cos\epsilon\,\sin2\psi\right]\,, \nonumber \\
h_{\times}(\Omega) & = & H \left[-\frac{1+\cos^{2}\epsilon}{2}\sin2\psi-i\,\cos\epsilon\,\cos2\psi\right]\,
\label{sig}.
\end{eqnarray}

\subsection{The signal matrix}

The GW response for a generator $Y^{(I)}=(p_{j}^{(I)},q_{j}^{(I)})$ is (see papers I and II), 

\begin{equation}
h^{(I)}(\Omega)=F_{+}^{(I)}(\Omega)h_{+}(\Omega)+F_{\times}^{(I)}(\Omega)h_{\times}(\Omega)\,,
\end{equation}
where,
\begin{eqnarray}
F_{+}^{(I)}(\Omega) & = & \sum_{i=1}^{3}\left(p_{i}^{(I)}F_{Vi;+}+q_{i}^{(I)}F_{Ui;+}\right) (\Omega )\, ,\nonumber \\
F_{\times}^{(I)}(\Omega) & = & \sum_{i=1}^{3}\left(p_{i}^{(I)}F_{Vi; \times}+q_{i}^{(I)}F_{Ui; \times}\right) (\Omega )\,, 
\end{eqnarray}
where we have expressed the response of the two polarisations in terms of the responses of the elementary data streams for 
each of the polarisations. Below we state the responses for the $+$ and $\times$ polarisations for the  beams 
$U^1$ and $V^1$ only; the responses for the remaining four beams $U^2, V^2, U^3, V^3$ are obtained from cyclic permutations:
\begin{eqnarray}
F_{U_1;+,\times} & = & \frac{e^{ i \Omega\left(\hat{w}\cdot\vec{r}_3 +
L_2\right)}} { 2 \left( 1 + \hat{w} \cdot \hat{n}_2 \right)} \ 
\left( 1 - e^{ - i \Omega L_2 \left( 1 + \hat{w} \cdot \hat{n}_2
\right)} \right) \xi_{2;+,\times} \, , \nonumber  \\
F_{V_1;+,\times} & = & -\frac{e^{ i \Omega\left(\hat{w}\cdot\vec{r}_2 +
L_3\right)}} { 2 \left( 1 - \hat{w} \cdot \hat{n}_3 \right)} \
\left( 1 - e^{ - i \Omega L_3 \left( 1 - \hat{w} \cdot \hat{n}_3
\right)} \right) \xi_{3;+,\times}  \, ,
\end{eqnarray} 
where,
\begin{equation}
\xi_{i;+} \ = \ (\hat{\theta}\cdot \hat{n}_i)^2 - (\hat{\phi}\cdot
\hat{n}_i)^2 , ~~~~~
\xi_{i;\times} \ = \ 2 (\hat{\theta}\cdot\hat{n}_i)(\hat{\phi}\cdot\hat{n}_i)
\, .
\end{equation}  
We define the signal covariance matrix $h^{(I)}_{(J)}$ in analogous fashion as the noise covariance matrix. It is 
given as follows: 
\begin{eqnarray}
h^{(I)}_{(J)} &=& h^{(I)} h_{(J)}^{*}, \nonumber \\
   &=& \left(F_{+}^{(I)}h_{+}+F_{\times}^{(I)}h_{\times}\right)\left(F_{+ (J)}h_{+}+ F_{\times (J)}h_{\times}\right)^{*}\,.
\end{eqnarray}
In general we may not have any knowledge of the polarisation of the GW binary source.  
We therefore average over the polarizations and assume that the direction of the orbital angular momentum of the 
binary is uniformly distributed over the sphere. The orientation of the binary (its orbital angular 
momentum vector) has been described in terms of the angles $\epsilon$ and $\psi$ in Eq.(\ref{sig}). Thus we carry 
out the averaging of $h^{(I)}_{(J)}$ over $(\epsilon, \psi )$ which results in an overall factor of $2/5$. 
The averaged matrix we denote by $\H^{(I)}_{(J)}$. In the Fourier domain it is given by,  

\begin{equation}
\H^{(I)}_{(J)}(\Omega) = H^2 \left( \frac{2}{5} \right) \left(F_{+}^{(I)} F_{+ (J)}^{*}
+ F_{\times}^{(I)} F_{\times (J)}^{*}\right) (\Omega) \, .
\end{equation}

The signal matrix so averaged has the following properties:

\begin{itemize} 

\item {$\H$ is the sum of outer products of two vectors: $F_{+}^{(I)}$ with its complex conjugate and 
$F_{\times}^{(I)}$ with its complex conjugate. Thus the natural basis for expressing $\H$ consists of the two vectors 
$F_{+}^{(I)}$ and $F_{\times}^{(I)}$. In the analysis that follows we will use this fact.}

\item {Because we average over the polarisations, $\H$ is constructed out of two vectors. Its rank is two, 
everywhere except on the  $\theta=\frac{\pi}{2}$ plane where it is one when $F_{\times}^{(I)}$ goes identically to zero. 
In paper II we had obtained a signal matrix of rank three because there we had averaged over the directions 
as well. While in \cite{PTL02}, since optimisation is performed first before averaging, the signal matrix is 
constituted from a single vector and thus has rank one.}  
 
\end{itemize}

\section{Optimizing SNR}
\label{SNR}

For a generic data combination $\alpha_{(I)}Y^{(I)}$ where $\a_{(I)}$ are polynomials in $E$, the SNR 
is given by:

\begin{equation}
SNR^{2}=\frac{\alpha_{(I)}\alpha^{(J) *}H^{(I)}_{(J)}}{\alpha_{(I)}\alpha^{(J) *}N^{(I)}_{(J)}}.
\label{genSNR}
\end{equation}
If the psd is given in units of Hz$^{-1}$ then the above equation yields the square of the SNR integrated over 
one second. Because one second is short compared to the various time-scales envisaged in the problem we call 
this SNR the {\it instantaneous} SNR. Since we are dealing with monochromatic sources, 
$E = e^{i \Omega L}$, the coefficients $\a_{(I)}$ reduce to 
just complex numbers. Also in the generating set $\{ Y^{(I)}\}$ the noise covariance matrix is diagonal 
with diagonal elements $n_{(I)}^2$. However, we find $n_{(1)} = n_{(2)}$ in our case, so that the first two 
eigenvectors correspond to the same eigenvalue. Thus, the denominator of Eq.(\ref{genSNR}) simplifies to a sum of 
squares ${\left|\alpha_{(1)}\right|^{2}n_{1}^{2}+\left|\alpha_{(2)}\right|^{2}n_{1}^{2}+
\left|\alpha_{(3)}\right|^{2}n_{3}^{2}}$ which we can set equal to unity because the SNR does not depend on the 
normalisation of the data combination. Moreover, it is convenient to define coefficients $\b$ which are scaled by the 
noise, $\beta_{(1)}=\alpha_{(1)}n_{1}, ~\beta_{(2)}=\alpha_{(2)}n_{2}, ~\beta_{(3)}=\alpha_{(3)}n_{3}$ so that  
the $\b_{(I)}$ satisfy,
\begin{equation}
\left|\beta_{(1)}\right|^{2}+\left|\beta_{(2)}\right|^{2}+\left|\beta_{(3)}\right|^{2}=1.
\label{cnstr}
\end{equation}
The expression for SNR simplifies to,
\begin{eqnarray}
SNR^2 & = & \alpha^{T}\cdot \H\cdot\alpha\ = \beta^{T}\cdot\rho\cdot\beta ,
\label{eq:condi}
\end{eqnarray}
where we define a SNR matrix $\rho$ by,
\begin{equation}
\rho^{(I)}_{(J)} = \frac{\H^{(I)}_{(J)}}{n_{(I)} n_{(J)}}.
\end{equation}
In the above equations the $\a, \b$ are construed of as $3 \times 1$ column matrices and $\H$ and $\rho$ as $3 \times 3$ square matrices.

The  extremisation of SNR is now carried out with the method of Lagrange multipliers because of the 
normalisation constraint on $\b$. This procedure yields an eigenvalue equation with the Lagrange multiplier 
appearing as an eigenvalue: 
\begin{equation}
\rho \cdot  \b = \lambda\beta \,.
\end{equation}
Since $\H$ has atmost rank 2, one eigenvalue is necessarily zero. We now proceed to compute the other two  
eigenvalues. The analysis is simplified if we go to the basis consisting of the two vectors $\vec{f_{+}}$ and 
$\vec{f_{\times}}$:
\begin{equation}
f^{(I)}_{+} = h_0 \frac{F^{(I)}_{+}}{n_{(I)}}, ~~~f^{(I)}_{\times} = h_0 \frac{F^{(I)}_{\times}}{n_{(I)}} ,
\end{equation}  
where $h_0 = (\sqrt{2/5}) H$ is the amplitude of the GW averaged over the polarisation states at frequency $\Omega$. 
We can then write the matrix $\rho$ as a tensor product in terms of these two vectors:
\begin{equation}
\rho = \f+ \otimes \f+^{*} + \fx \otimes \fx^{*}.
\end{equation}
In general the vectors $\f+$ and $\fx$ are not orthogonal.

The action of the 
matrix $\rho$ on any vector $\vec v$ is  given by,
\begin{equation}
\rho.\vec v \ = \ (\vec f_+ \otimes \vec f_+^* + \vec f_\times \otimes \vec
f_\times^*)\cdot \vec v  \ = \  (\vec f_+^*\cdot\vec v) \vec f_+ +
(\vec f_\times^*\cdot\vec v) \vec f_\times \,.
\end{equation}
For the eigenvalue problem, we have the eigenvalue equation,
\begin{equation}
 (\vec f_+^*\cdot \vec v) \vec f_+ +(\vec f_\times^*\cdot \vec v) \vec f_\times \ = \
\lambda \vec v\,.
\end{equation}
Expressing the eigenvector $\vec{v}$  as a linear combination of $\vec f_+$ and
$\vec f_\times$,
$\vec v \ = \ c_+ \vec f_+ + c_\times \vec f_\times\,$,
and from the linear independence of $\f+$ and $\fx$ (they are linearly independent
in general) we obtain the system of
equations for $c_+$ and $c_{\times}$ as:
\begin{equation}
\left\{ 
\begin{array}{l}
(| \vec f_+ |^2 - \lambda) c_+ + ( \vec f_+^*\cdot \vec f_\times) c_\times \ = \ 0 \\
(\vec f_+ \cdot  \vec f_\times^*) c_+ + (| \vec f_\times |^2 - \lambda) c_\times \ =
\ 0 \, .
\end{array}
\right. \,  
\label{eqn:eigncnd}
\end{equation}
 Setting the determinant of this system to zero,
we obtain eigenvalue equation:

\begin{equation}
\lambda^2 - (|\f+ |^2 + |\fx |^2 ) \lambda  + |\f+ \times \fx |^2 = 0 \,,
\end{equation}
which can be solved to yield the eigenvalues:
\begin{equation}
\lambda_{\pm} = {\frac{1}{2}} ( |\f+ |^2 + |\fx |^2  \pm \sqrt{\Delta}),
\end{equation}
where, 
\begin{eqnarray}
\Delta &=& (|\f+ |^2 + |\fx|^2)^2 - 4 |\f+ \times \fx|^2 \nonumber \\
       &=& (|\f+ |^2 - |\fx |^2 )^2 + 4 |\f+ \cdot \fx^{*} |^2. 
\end{eqnarray} 
In the low frequency limit as we shall find 
$\f+ \cdot \fx^{*} \approx 0$ so that the positive square root of $\Delta$ is just $|\f+ |^2 - |\fx|^2$ and the 
eigenvalues are just $|\f+|^2$ and $|\fx|^2$ (infact the eigenvalue equation factorises into linear factors) 
with $\f+$ and $\fx$ as eigenvectors respectively. This can also be directly inferred from the structure of $\rho$. 
So it may be appropriate to call $\lambda_{-}$ as $\lambda_{\times}$ even in the general case. In the general case 
the eigenvectors can be easily determined from Eq.(\ref{eqn:eigncnd}):
\begin{eqnarray}
\vec{v}_{+} &=& (\lambda_{+} - | \fx|^2) \f+ + (\fx^{*} \cdot \f+) \fx \,, \nonumber \\
\vec{v}_{\times} &=& (\fx \cdot \f+^{*}) \f+ + (\lambda_{\times} - |\f+ |^2) \fx \,,
\end{eqnarray}   
where $\vec{v}_{+, \times}$ are eigenvectors belonging to the eigenvalues $\lambda_{+, \times}$ respectively. 
Since $\rho$ is Hermitian the eigenvectors are orthogonal, $\vec{v}_{+}\cdot \vec{v}_{\times}^{*} = 0$ as can be verified 
also directly from the expressions of the eigenvectors. These eigenvectors are not normalised. 

The eigenvalues $\lambda_{+, \times}$ are the squares of the instantaneous SNRs for the two data combinations  
described by the two corresponding eigenvectors. The data combinations are $\vec{v}_{+ (I)} Y^{(I)}$ and
$\vec{v}_{\times (I)} Y^{(I)}$, which we will call eigen-combinations or alternatively eigen-observables, and 
which give the instantaneous SNRs: SNR$_{+, \times}^2 \equiv \lambda_{+, \times}$ respectively. 
For a given direction $(\theta, \phi)$ in the LISA 
frame,  SNR$_{+}$ is the maximum instantaneous SNR among all data combinations. While  SNR$_{\times}$ is the 
minimum instantaneous SNR among data combinations that are linear combinations of 
$\vec{v}_{+}$ and $\vec{v}_{\times}$; those which lie in the `plane' of $\vec{v}_{+}$ and $\vec{v}_{\times}$.
However, SNR$_{\times}$ is not zero in general and therefore not the absolute minimum; the absolute minimum SNR 
is zero corresponding to the third eigenvalue which is zero. 
Moreover, the eigenvectors are orthogonal, which means they yield statistically independent observables. 
Thus these observables can be combined in quadratures to form a network observable with  
instantaneous SNR$_{\rm network}$ given by:
\begin{eqnarray}
{\rm SNR}_{\rm network}^2 &=& {\rm SNR}_{+}^2 + {\rm SNR}_{\times}^2, \nonumber \\ 
                          &=& \lambda_{+} + \lambda_{\times} = |\f+ |^2 + |\fx |^2 . 
\end{eqnarray}   
The third eigenvector is $\f+^{*} \times \fx^{*}$; it is orthogonal
to $\f+$ and $\fx$ with eigenvalue zero \footnote{While defining the vector cross product of 
two complex vectors, we require its scalar product with both vectors
to vanish.
Since the scalar product is defined 
by taking the complex conjugate of the second vector, complex conjugates appear in the cross product.}. 
This means that the data combination corresponding to this vector gives zero response in that particular direction, 
which may be important if one wishes to `switch off' the GW coming from that direction.  

\section{Tracking a GW source with LISA}
\label{track}

In general the amplitude of the GW source will be small and it would be necessary to track or follow the 
source for a considerable period of time in order to accumulate an adequate SNR. This period of time could 
range from few days to a year or even years - the life of the LISA experiment. The LISA configuration 
performs a complex motion due to which the source will appear to move in the LISA frame even if it is 
stationary in the barycentric frame. For the purposes of this analysis, we will take the observation time 
to be a year and integrate the SNR for this period of time. It is possible that some GW sources may be 
sufficiently powerful that SNR integration for an year is not required. In that case it is easily possible 
to limit the integration to the required period of observation and obtain useful results. We will also 
assume that the source is stationary in the barycentric frame, that is, its direction remains constant during the 
observation period. In this section we will present the transformations connecting the barycentric frame and the 
LISA frame and hence obtain the apparent motion in the LISA frame for a GW source fixed in the barycentric frame. 

The LISA constellation trails the Earth in its orbit by $20^{\circ}$ around the sun. The plane of LISA makes an angle of 
$60^{\circ}$ with the plane of the ecliptic and the LISA triangle rotates in its own plane completing one rotation in 
a year. We describe the barycentric frame by the Cartesian coordinates $\{x_B, y_B, z_B \}$. The $x_B-y_B$ plane 
coincides with the orbital plane of LISA. The $z_B$-axis is orthogonal to this plane forming a right-handed 
coordinate system. The transformation from the barycentric frame to LISA frame is given as follows \cite{LISArep}: A vector 
${\bf r_B} = (x_B, y_B, z_B)^T$ is transformed to ${\bf r_L} = (x_L, y_L, z_L)^T$ by the matrix ${\cal R}$ as, 
\begin{equation}
{\bf r_L} = {\cal R}\cdot {\bf r_B},
\end{equation}
where ${\cal R}$ is a product of the three matrices containing Euler angles:
\begin{equation}
{\cal{R}}  =  \mathcal{C}\cdot\mathcal{B}\cdot\mathcal{A}\,,
\end{equation}
where the matrices ${\cal A}$, ${\cal B}$ and ${\cal C}$ are given by,
\begin{eqnarray}
\mathcal{A} & = & \left(\begin{array}{ccc}
\cos\psi_{a} & \sin\psi_{a} & 0\\
-\sin\psi_{a} & \cos\psi_{a} & 0\\
0 & 0 & 1\end{array}\right) \, , \nonumber \\
\mathcal{B} & = & \left(\begin{array}{ccc}
1 & 0 & 0\\
0 & \frac{1}{2} & \frac{\sqrt{3}}{2}\\
0 & -\frac{\sqrt{3}}{2} & \frac{1}{2}\end{array}\right) \, , \nonumber \\
\mathcal{C} & = & \left(\begin{array}{ccc}
\cos\psi_{c} & \sin\psi_{c} & 0\\
-\sin\psi_{c} & \cos\psi_{c} & 0\\
0 & 0 & 1\end{array}\right) \, .
\end{eqnarray}
Here $\psi_{a}=\omega t+\alpha_{0}$, $\psi_{c}=-\omega t+\beta_{0}$ and $\omega={2\pi}/{T_{\odot}}$.
$T_{\odot}$ is the orbital period of LISA which we take to be of one year duration. 
The $\alpha_0$ and $\beta_0$ are constants fixing initial conditions when the observation begins. 
The matrix ${\cal R}$ is time-dependent and is a product of two time-dependent matrices.

\begin{figure}  
\caption{Apparent position of the source in the sky as seen from LISA frame for $(\theta_B = 0,\,\phi_B = 0\,)$.  
The track of the source for a period of one year is shown on the unit sphere
in the LISA frame.
}

\includegraphics[width=.5\textwidth]{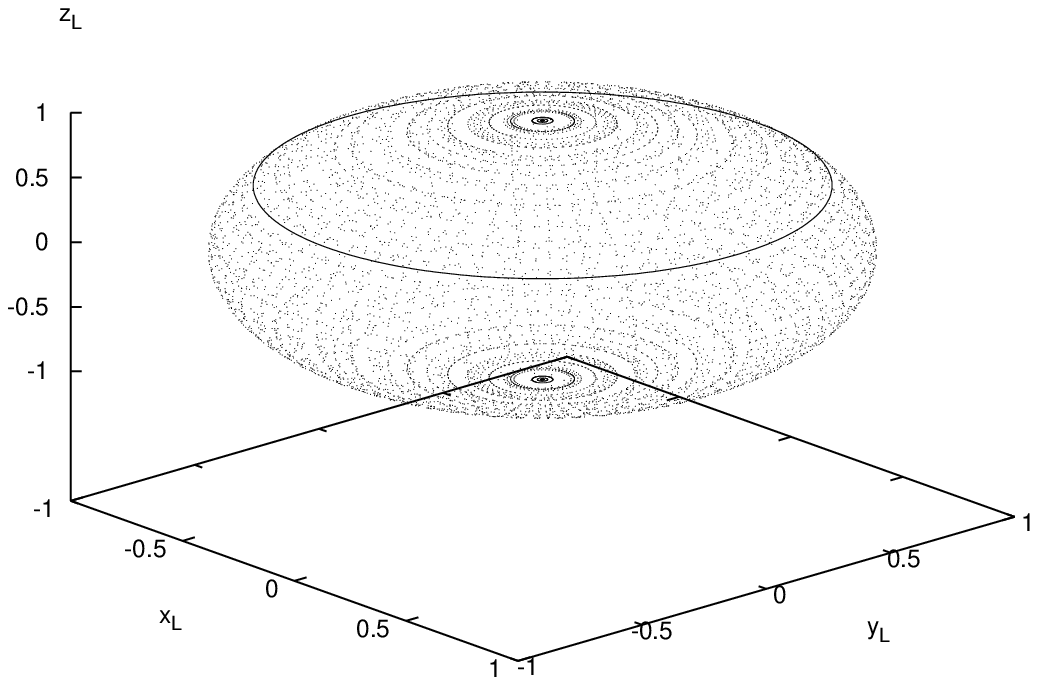}
\label{trk1}

\caption{Apparent position of the source in the sky as seen from LISA frame for  $(\theta_B=\frac{\pi}{2},\,\phi_B=0\,)$.
The track of the source for a period of one year is shown on the unit sphere
in the LISA frame.
}

\includegraphics[width=.5\textwidth]{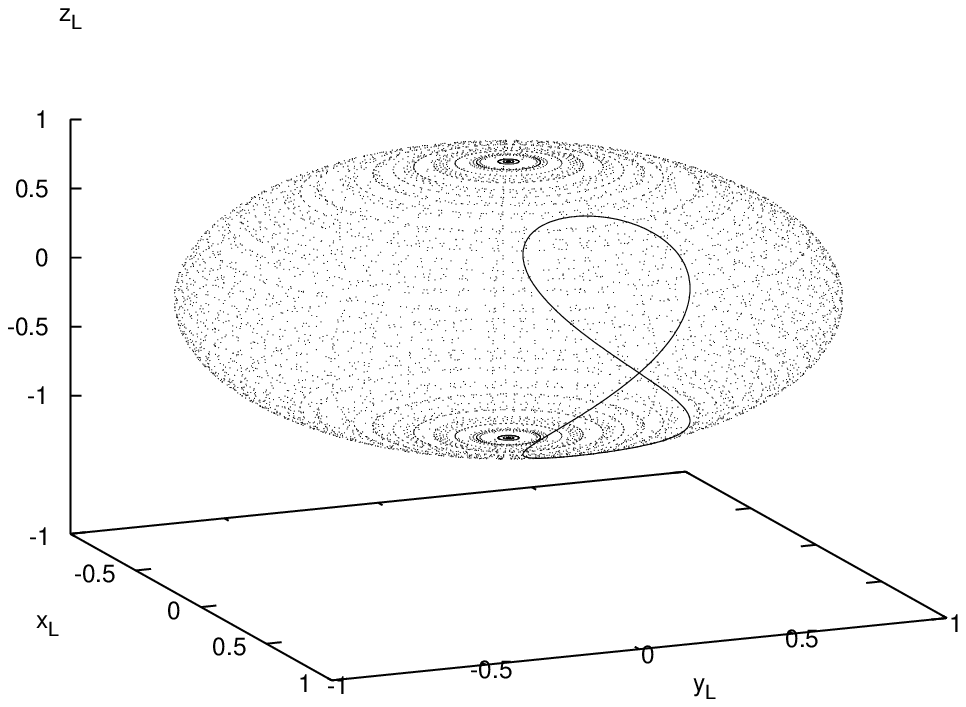}
\label{apptrk}
\end{figure}

The unit vector $\hat{w}_{B}$ of the source direction described by the angles $(\theta_B, \phi_B)$ in the barycentric 
frame is given by $\hat{w}_B = (\sin\theta_{B}\cos\phi_{B}, \sin\theta_{B}\sin\phi_{B}, \cos\theta_{B})$. The 
corresponding vector $\hat{w}_L$ in the LISA frame is then, $\hat{w}_{L}={\cal R}(t)\cdot\hat{w}_{B}$. From this vector we can 
explicitly work out the angles $\left(\theta_{L},\,\phi_{L}\,\right)$:

\begin{eqnarray}
\theta_{L} & = & \cos^{-1}\left(\frac{1}{2}\cos\theta_{B}
               -\frac{\sqrt{3}}{2}\sin\theta_{B}\sin\left(\phi_{B}-\psi_{a}\right)\right), \nonumber \\
\phi_{L} & = & \tan^{-1}\left(\frac{w_{L}^{2}}{w_{L}^{1}}\right),
\label{trns}
\end{eqnarray}
where,
\begin{eqnarray}
w_L^1 &=& \cos\psi_{c}\sin\theta_{B}\cos\left(\phi_{B}-\psi_{a}\right) \nonumber \\
&& +\frac{1}{2}\sin\psi_{c}\sin\theta_{B}\sin\left(\phi_{B}-\psi_{a}\right) \nonumber \\
&& +\frac{\sqrt{3}}{2}\sin\psi_{c}\cos\theta_{B}, \nonumber \\
w_L^2 &=& -\sin\psi_{c}\sin\theta_{B}\cos\left(\phi_{B}-\psi_{a}\right) \nonumber \\
&& +\frac{1}{2}\cos\psi_{c}\sin\theta_{B}\sin\left(\phi_{B}-\psi_{a}\right) \nonumber \\
&& +\frac{\sqrt{3}}{2}\cos\psi_{c}\cos\theta_{B}.
\end{eqnarray}
For a fixed source direction $(\theta_B, \phi_B)$ in the barycentric frame, the apparent source direction 
depends on time in the LISA frame; $\theta_L, \phi_L$ are functions of time; the source appears to move 
in the LISA frame. Thus a given data combination even if it is optimal initially, will not continue to remain optimal 
subsequently. Our strategy is then to {\it switch} the data combinations continuously so that the SNR remains 
optimal at all times. In this way we can optimally track the source and accumulate maximum SNR. The following figures  
show the apparent trajectory of the source for one year in the LISA frame when (i) the source lies at the pole of the 
barycentric frame - it is just a circle in the LISA frame: $\theta_L = \pi/3$ (See Fig.\ref{trk1}) and 
(ii) the source lies in plane of the LISA orbit - a figure of 8 is described by the source (See Fig.\ref{apptrk}). 

\section{Optimal sensitivities}
\label{sens}

For optimal tracking of the source, we switch the data combinations so that the SNRs yielded 
are optimal at all times. Three SNRs are of interest; $SNR_{+, \times}$ and SNR$_{\rm network}$, that is, the two eigenvalues 
and their sum. The eigenvalues are functions of $\theta_L, \phi_L$ which are in turn functions of $t$ as the GW source describes 
its apparent motion in the LISA frame. The three integrated SNRs are given by:
\begin{eqnarray}
{\rm SNR}_{+, \times}^2 (\hat{w}_B) &=& \int_{0}^{T} \lambda_{+, \times} (\theta_L (t; \hat{w}_B), \phi_L (t;\hat{w}_B )) dt, \nonumber \\
{\rm SNR}_{\rm network}^2 (\hat{w}_B) &=& {\rm SNR}_{+}^2 (\hat{w}_B) + {\rm SNR}_{\times}^2 (\hat{w}_B), 
\end{eqnarray}
where we have taken the total observation time to be $T$ and the initial time of observation to be zero. If we integrate 
for a complete year, we can set the initial observation time to be zero without loss of generality and at the same time 
set the constants $\alpha_0, \beta_0$ to be zero. However, if we make observations for times that are 
not integral number of years, the constants $\alpha_0, \beta_0$ must be chosen appropriately in the transformation 
matrix ${\cal R}$.

An important point is that, since LISA moves in an heliocentric orbit, a Doppler phase depending on the position vector of the 
LISA  projected on to the direction to the source $\hat{w}_B$ and the GW frequency will be added to the phase of the GW signal. 
This Doppler phase will be a function of time which will be added 
to the monochromatic part of the phase $\Omega t$ of the signal. We assume in this analysis that this phase has been accounted 
for when integrating the signal. One way is to `stretch' the time-coordinate so that the signal appears monochromatic 
(a technique well-known to radio astronomers)\cite{ref:Schz,ref:Schz1}.    
  
\subsection{The Low Frequency Limit}

An important case arises when we consider GW of low frequencies, say below 3 mHz. For this case it is possible to obtain 
analytical expressions for the optimal SNRs. A large fraction of the sources for LISA fall into this category, for example, 
massive/supermassive blackhole binaries, several galactic and extragalactic binaries which contribute to the `confusion noise'.  

We need to compute the $\f+$ and $\fx$ in order to compute the SNR matrix $\rho$ and then obtain its eigenvalues. Integrating 
the eigenvalues for the period of observation will yield the relevant SNRs. To this end, we expand
$F_{Ui;+,\times}$ and $F_{Vi;+,\times}$ to the lowest order in the dimensionless frequency $\Omega L$:

\begin{eqnarray}
F_{U1;+,\times} & = & i\frac{\Omega L}{2}\left(1+i\Omega L\,\tau_{2}\right)\xi_{2;+,\times}, \nonumber \\
F_{V1;+,\times} & = & -i\frac{\Omega L}{2}\left(1+i\Omega L\,\tau_{3}\right)\xi_{3;+\times},
\end{eqnarray}
where,
\begin{eqnarray}
\tau_{m} & = & \frac{1}{2}\left(1-\frac{\hat{w}\cdot\hat{r}_{m}}{\sqrt{3}}\right)\,,\nonumber \\
\xi_{m;+} & = & \frac{\left(1+\cos^{2}\theta\right)}{2}\cos\left(2\phi- (2m -1)\frac{\pi}{3}\right)-
\frac{1}{2}\sin^{2}\theta \,, \nonumber \\
\xi_{m;\times} & = & -\cos\theta\sin\left(2\phi- (2m - 1)\frac{\pi}{3}\right)\,, 
\end{eqnarray}
where, $\hat{r}_{m}$ is the unit vector in the direction of $m$-th spacecraft and $m = 1, 2, 3$ and the angles $\theta, \phi$ 
refer to the LISA frame (as before we have dropped the subscript `$L$'). The transfer functions 
for the four other elementary data streams are obtained by cyclic permutations. In order to get the 
$\f+, \fx$ we must operate on the $U^i, V^i$ with the polynomial operators $p_i^{(I)}, q_i^{(I)}$ given 
in Eq.(\ref{basis}) and then scale them by the averaged signal divided by the noise. Thus, we obtain after some algebra, 
\begin{eqnarray}
\f+ & = & \rho_0 \frac{(1 + \cos^2 \theta)}{2} ( - \sin \Phi, \cos \Phi, 0 ), \nonumber \\
\fx & = & - \rho_0 \cos \theta (\cos \Phi, \sin \Phi, 0), 
\end{eqnarray}
where, 
\begin{equation}
\rho_0 =  \frac{3}{\sqrt{5}} \frac{H}{n_1} (\Omega L)^2 ,~~~~\Phi = 2 \phi + \frac{\pi}{3}.
\end{equation}
The  vectors $\f+$ and $\fx$ are expressed in the $Y^{(I)}$ basis and
in this basis 
they have real components. We observe the 
following properties of the vectors: 
\begin{itemize}
\item{
The vectors $\f+$ and $\fx$ lie in the $(Y^{(1)}, Y^{(2)})$ plane and the $Y^{(3)}$ component is 
zero for both vectors.  This can be
understood if we recall that $Y^{(3)}$ is proportional to symmetric Sagnac
combination \cite{SNV03} which is insensitive to GW at low frequency.}
\item{The apparent motion of a GW source in the LISA frame can be optimally and continuously tracked by $\hat{f}_+$ and
$\hat{f}_\times$ by rotating the pair $(\hat{f}_+, \hat{f}_{\times})$ by $\Phi$ at each instant of time as follows:
\begin{equation}
\left(\begin{array}{c}
\hat{f}_\times \\
\hat{f}_+ \end{array}\right) =
\left(\begin{array}{cc}
\cos\Phi & \sin\Phi\\
-\sin\Phi & \cos\Phi\end{array}\right) \, \left(\begin{array}{c}
Y^{(1)}\\
Y^{(2)} \end{array}\right) \,. \
\end{equation} 
Here $\hat{f}_+$ and $\hat{f}_\times$ are unit vectors in the directions of $\f+$ and $\fx$ respectively. 
Thus, optimally tracking a source amounts to orienting 
the data combinations along
$\f+$ and $\fx$.}
\item{Moreover, the vectors are orthogonal: $\f+ \cdot  \fx^{*} = 0$ in 
this low frequency limit. The orthogonality implies that these eigen-observables
$\f+$ and $\fx$ have zero response to $\times$ and $+$ polarizations of GW
  respectively.  We use this fact to estimate the polarisation
  angles; namely $(\epsilon,\psi)$ at the end of this section.} 
\end{itemize}

 The eigenvalues of the eigen-vectors $\f+$ and $\fx$ are just 
$|\f+|^2$ and $|\fx|^2$ and are explicitly given by,
\begin{eqnarray}
\lambda_{+} & = &  \rho_0^2 \left (\frac{1+\cos^{2}\theta}{2} \right)^2 \,, \nonumber \\
\lambda_{\times} & = &  \rho_0^2 \cos^{2}\theta\,.
\label{eigSNR}
\end{eqnarray}
The eigenvalues are the squares of the instantaneous SNRs. 
We notice that $\rho_0$ is the maximum instantaneous  SNR obtained when the 
source lies at the poles $\theta = 0$ or $\pi$ in the LISA frame. If the source is observed over a period of an year, 
the eigenvalues must be integrated over this length of time. We notice that the eigenvalues do not depend on $\phi$. 
Thus our next task of integrating the SNR becomes somewhat simplified.  

\subsubsection{The integrated SNR}
\begin{figure}
\caption{Instantaneous $SNR_{+, \times}$ and SNR$_{\rm network}$ as functions of time for the source direction 
$(\theta_B=\frac{\pi}{2},\,\phi_B=0\,)$ at the GW frequency of $f=1$ mHz in units of $\rho_0$.}

\includegraphics[width=0.5\textwidth]{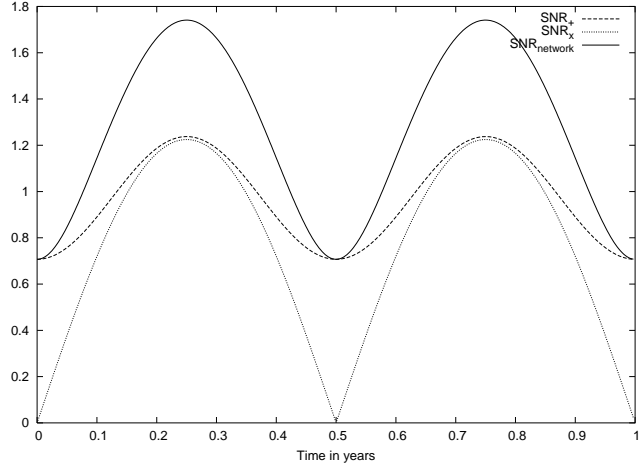}
\label{SNRtime}
\end{figure}

For a given source direction $(\theta_B, \phi_B)$ the corresponding
track $(\theta_L(t), \phi_L(t))$  of the 
source in LISA frame is given by Eq.(\ref{trns}). We may substitute 
these values into the integrals for the SNRs and the integrals yield simple analytical expressions for the SNRs if the 
integration is over an integral number of years. The instantaneous SNRs however change with time. 
In the Fig.\ref{SNRtime} we show how the SNRs change with time as LISA orbits the sun during the course of a year.  
The various SNRs  are shown for a source lying in the ecliptic plane, $(\theta_B = \pi/2, \phi_B = 0)$ for the GW frequency 
of 1 mHz. We have chosen this direction because the SNRs show considerable variations during the course of a year.  

\begin{figure}
\caption{The functions $g_{+, \times}$ and $g_{\rm network}$ as functions of $\theta_B$}

\includegraphics[width=0.5\textwidth]{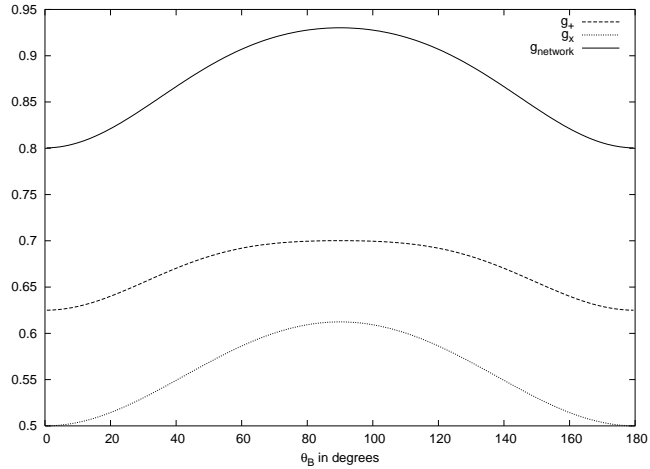}
\label{intSNR}
\end{figure}

Integration over a period of $T = T_{\odot}$ leads to the following results:
\begin{eqnarray}\label{eq:SNR+x}
SNR_{+}(\theta_B, \phi_B) & = & {\rm SNR}_0 ~ g_{+} (\cos \theta_B) \,, \nonumber \\
SNR_{\times} (\theta_B, \phi_B)& = & {\rm SNR}_0 ~ g_{\times} (\cos \theta_B)\, , 
\end{eqnarray}
where, 
\begin{equation}
{\rm SNR_0} = \rho_0 \sqrt{T_{\odot}} = \frac{3}{\sqrt{5}} \frac{H}{n_1} (\Omega L)^2 \sqrt{T_{\odot}} \,,
\label{SNR0}
\end{equation}
and  
\begin{eqnarray}
g_{+}^2 (x) &=& \frac{1}{T_{\odot}} \int_0^{T_{\odot}} \left( \frac{1 + \cos^2 \theta_L}{2} \right)^2 dt \nonumber \\
            &=& \frac{1}{4} \left (1 + \frac{x^2}{4}\right)^2\nonumber  \\ 
             && + \frac{3}{16} (1 - x^2) \left(1 + \frac{3}{4} x^2 + \frac{9}{32} \left(1 - x^2 \right) \right),\nonumber  \\
g_{\times}^2 (x) &=& \frac{1}{T_{\odot}} \int_0^{T_{\odot}} \cos^2 \theta_L dt\nonumber  \\
               &=&  \frac{1}{4} \left[x^2 + \frac{3}{2} \left(1 - x^2 \right)\right].
\end{eqnarray} 
We have purposely not `simplified' the formulae in powers of $x^2$ because in this form it is easy to see the 
limits $x = \pm 1, 0$ corresponding to $\theta_B = 0, \pi, \pi/2$ respectively. Infact since only $x^2$ occurs in the 
expressions of $g_{+, \times}$ there is symmetry about the ecliptic plane. The network SNR is just the root mean square 
of the two SNRs:
\begin{equation}
{\rm SNR}_{\rm network} (\theta_B, \phi_B) = {\rm SNR}_0 ~ g_{\rm network} (\cos \theta_B)\, ,
\end{equation}
where, $g_{\rm network}^2 (x) = g_{+}^2 (x) + g_{\times}^2 (x)$. In Fig.\ref{intSNR} we plot the functions 
$g_{+, \times}$ and $g_{\rm network}$ as functions of $\theta_B$ between $0^o \leq \theta_B \leq 180^o$. The factors $g$ are of 
the order of unity and the SNR$_0$ gives essentially the integrated SNR of a GW source. 
Recall that this is an SNR averaged over the polarisations.

We observe from Fig.\ref{intSNR} that the maximum integrated SNR is obtained for sources lying in the ecliptic plane 
$(\theta_B = 90^{\circ})$. This can be readily explained from Fig.\ref{apptrk} where the trajectory of such a source is 
plotted. One observes that a large fraction 
of the orbit in the LISA frame is {\it away} from the LISA plane $(\theta_L = 90^{\circ})$. As seen from Eq.(\ref{eigSNR}) 
the sensitivity of LISA increases as we go away from the plane of LISA, that is, towards the poles.
We also observe that the variations in the three curves are small; among these the network SNR shows the highest variation of 
$\approx 15 \%$. This shows that LISA has a more or less uniform average response over the year as it moves in its orbit.    
The above formulae are also more generally valid if the integration time $T$ is taken to be in integral multiples of $T_{\odot}$; 
then the $T_{\odot}$ in Eq.(\ref{SNR0}) should be replaced by $T$.  

We compute the SNRs of six binaries in our galaxy which give high average SNR (averaged over polarisations) if we 
integrate the SNR optimally over the period of a year. The following table lists six binary systems which give the 
network SNRs ranging from about 3 to over 100. The SNRs have been computed assuming circular orbits for the binaries. 
The information about the binaries has been obtained from \cite{ctlg}. 
We observe that the binary masses are small; 
$\sim .5 M_{\odot}$ and companion mass $\sim .02 M_{\odot}$. The reason for  
such small masses is that the orbital periods of the binaries must be short, in this case ranging from about .01 to .03 of a day. 
The GW quadrupole 
frequency $f_{gw}$ is related to the orbital period by the equation:
\begin{equation}
f_{gw} = \frac{2}{P_{orb}} = 2.3 \left( \frac{P_{orb}}{.01 {\rm day}} \right)^{-1} {\rm mHz}.
\end{equation}
Thus the sources radiate GW at frequencies $\sim$ 1 or 2 mHz, the band in which LISA has maximum sensitivity. This yields 
the high SNRs. They are also close by; $R \sim 100$ or 200 pc which tends to increase the raw gravitational wave amplitude $H$. 
Note that the SNRs listed in the table are average and actual observations can yield different values depending on the orientation of the binary orbit. 
Also the observables used 
are optimal in the average sense. If the orientation of the binary is known then in general a better observable can be found.

\vspace{.2in}
\begin{center}
\begin{table}
\caption{In this table we list six binaries in our galaxy whose parameters are known. For these we compute the optimum SNRs 
where the optimisation has been performed after averaging over the polarisations.} 
\begin{tabular}{|c|c|c|c|c|c|c|c|c|c|}
\hline 
Name&
$P_{orb}$ in days&
$m_{1}$&
$m_{2}$&
$\log(H)$&
$\theta_{B}^{\circ}$&
$R$ in pc&
$SNR_{+}$&
$SNR_{\times}$&
$SNR_{network}$\tabularnewline
\hline
\hline 
AM CVn&
0.011907&
0.5&
0.03&
-21.4&
124.46&
100&
89.9&
75.6&
117.4\tabularnewline
\hline 
CP Eri&
0.019950&
0.6&
0.02&
-22.0&
116.43&
200&
8.9&
7.4&
11.4
\tabularnewline
\hline 
CR Boo&
0.017029&
0.6&
0.02&
-21.7&
72.103&
100&
26.3&
22.7&
34.7
\tabularnewline
\hline 
GP Com&
0.032310&
0.5&
0.02&
-22.2&
66.997&
200&
2.2&
1.8&
2.8
\tabularnewline
\hline 
HP Lib&
0.012950&
0.6&
0.03&
-21.4&
85.040&
100&
77.6&
67.8&
103.0
\tabularnewline
\hline 
V803 Cen&
0.018650&
0.6&
0.02&
-21.7&
120.32&
100&
20.6&
17.5&
27.0\tabularnewline
\hline
\end{tabular}
\label{bin}
\end{table}
\end{center}

\subsubsection{Estimation of inclination angle of the orbital plane}
We showed in previous sections that if we do not have any prior
information of the inclination angle $\epsilon$ of the binary orbit and the
the angle $\psi$, then tracking the source with $\hat{f}_+$ and $\hat{f}_\times$ are optimal in the
average sense. Typically, the orbital inclination is difficult to
estimate from other astrophysical observational means. However, for binaries
with known masses and distances (e.g. binaries in table-{\ref{bin}}), we can 
estimate $(\epsilon,\psi)$ from the output of
eigen-observables $\f+$ and $\fx$.

The integrated SNRs over the period of one year of $\f+$ and $\fx$, {\it without averaging over polarisations} are given by
\begin{eqnarray}  \label{eq:estpol}
		SNR_{+}(\theta_B, \phi_B) & = & \sqrt{\frac{5}{2}}~{\rm SNR}_0 ~ g_{+} (\cos \theta_B)~ a_+\,, \nonumber \\
		SNR_{\times} (\theta_B, \phi_B) & =& \sqrt{\frac{5}{2}}~ {\rm SNR}_0 ~
g_{\times} (\cos \theta_B)~ a_\times,
\label{est}
\end{eqnarray}
where $a_+ = |h_+ (\Omega) /H|$ and $a_\times = |h_\times (\Omega) /
H|$ (see Eq. (\ref{sig})). We note that here the factor $\sqrt{5/2}$ appears, since
there is no averaging over the polarisations. The $a_{+, \times}$ can be estimated, if the SNRs appearing 
on the LHS of Eq. (\ref{est}) can be measured. Further, straightforward algebra shows that 
\begin{equation}
a_+^2 + a_\times^2 = \left({1+\cos^2\epsilon} \over 2 \right)^2 +
\cos^2\epsilon .
\end{equation}
From the above equation, we can estimate $\epsilon$. Substituting back in $a_+$, one can also estimate $\psi$ if needed.
This exercise can be carried out for the binaries listed in the table \ref{bin}.

\subsection{The general case}

\begin{figure}
\caption{Sensitivity curves for the observables: Michelson, Switched Michelson, $\vec{v}_+$ and network for the 
source direction $\theta_B = 90^{\circ}, \phi_B = 0^{\circ}$.}
\includegraphics[width=0.5\textwidth]{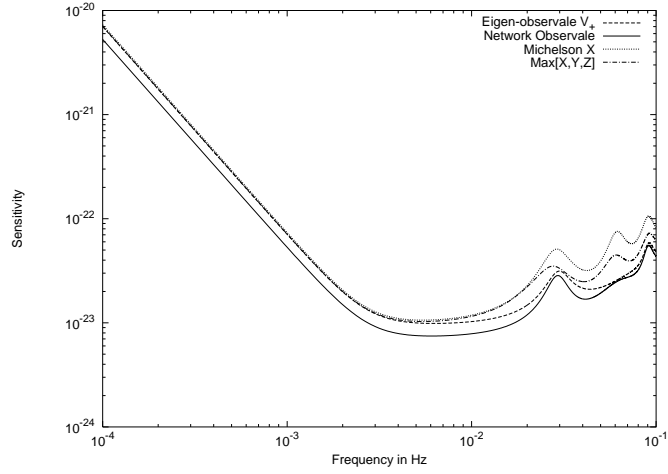}
\label{fig:sens1}
\end{figure}

\begin{figure}
\caption{Ratios of the sensitivities of the observables network, $\vec{v}_{+, \times}$ with $\X$ for the source 
direction $\theta_B = 90^{\circ}, \phi_B = 0^{\circ}$.}
\includegraphics[width=0.5\textwidth]{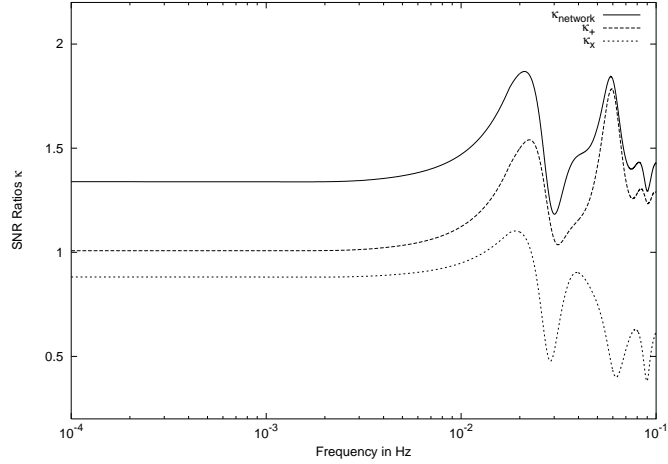}
\label{fig:SNRratio}
\end{figure}

In this section we relax the condition of dealing only with low frequencies and consider the entire band-width of LISA. We 
compare the sensitivities for LISA obtained by using the optimum SNR with the Michelson data combination usually denoted by 
$X$. As a polynomial vector $(p_i, q_i)$ it is given by:
\begin{equation}
X = (1 - E_2^2 , 0, E_2 (E_3^2 - 1), 1 - E_3^2, E_3 (E_2^2 - 1), 0).
\end{equation}
When we set all the $E_i$ to be equal, the factor $(1 - E^2)$ factors out and one is left with a simple polynomial vector 
$(1, 0, -E, 1, -E, 0)$. This combination has been used to plot the standard LISA sensitivity curve. 
Two other Michelson observables are obtained by cyclic permutations called $Y$ and $Z$ in the literature \cite{AET99}. 
In this section we will compare the sensitivities obtained by using the eigen-combination $\vec{v}_+$ and the network observable 
with the Michelson combination $X$ and the observable $\X$ obtained by `switching' the Michelson's $X, Y, Z$ optimally, that is,
$SNR_{\X} = \max (SNR_X, SNR_Y, SNR_Z)$.
Extending the definition of sensitivity of a data combination from earlier literature  \cite{LISArep}
to an observable, we define the sensitivity of an observable $W$ as,
\begin{equation}
    {\rm Sensitivity}_W(f) = \frac{5}{SNR_W(f)}\, ,
    \label{eq:intsen}
\end{equation}
where $SNR_W(f)$ is the integrated SNR over a observation period $T$. 
 The number $5$ has been chosen following earlier literature.
Eq. (\ref{eq:intsen}) for a fixed data combination $W_0$ reduces to the 
standard one in  literature:
\begin{equation}
    {\rm Sensitivity}_{W_0}(f) = 5 \frac{\sqrt{S_{W_0} (f) B}}
    {|h_{W_0}|}  \, ,
\end{equation}
where $B = 1/T$. As before we take $T = T_{\odot}$ for plotting the sensitivity curves in the figures below. 
The results of our findings are 
displayed in the plots. Fig.\ref{fig:sens1} displays the sensitivity curves for the observables (a) Michelson - $X$  (dotted  curve),  
(b) Switched Michelson $\X$ (dash-dotted curve), (c) Eigen-combination $\vec{v}_+$ (dashed curve) and (d) network 
observable (solid curve).
 We observe that the sensitivity over the band-width of LISA increases as we go from (a) to (d). 
We observe that the $\X$ does not do much better than $X$. This is because for the source direction chosen $X$ is reasonably 
well oriented and there is not much switching to $Y$ and $Z$ combinations. However, the network and $\vec{v}_+$ observables 
show significant improvement in sensitivity over both $X$ and $\X$. The sensitivity curves (except $X$) do not show 
much variations for other source directions and the plots are similar. 
The quantitative comparison of sensitivities is shown in Fig.\ref{fig:SNRratio}. 

\begin{figure}
\caption{The sensitivity curves for the network observable for $\theta_B = 0^{\circ}, 30^{\circ}, 60^{\circ}, 90^{\circ}$}
\includegraphics[width=0.5\textwidth]{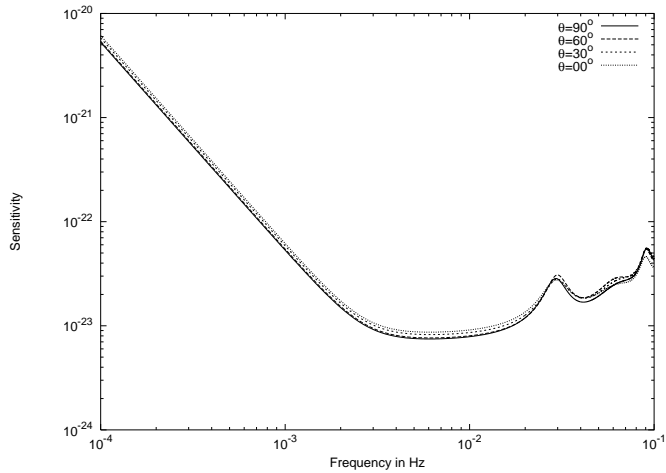}
\label{fig:sens-net}
\end{figure}

Here we have compared the network, and the eigen-combinations $\vec{v}_{+, \times}$ with $\X$. Defining:
\begin{equation}   
\kappa_a (f) = \frac{{\rm SNR}_a(f)}{{\rm SNR}_{\X}(f)} \, ,
\end{equation}
where the subscript $a$ stands for network or $+, \times$ and ${\rm SNR}_{\X}$ the SNR of the observable $\X$, 
we plot these ratios of sensitivities over the LISA band-width. We notice from the $\kappa_{\rm network}$ curve that the 
improvement in sensitivity for the network observable is about 34$\%$ at low frequencies and rises to nearly 90 $\%$ at 
about 20 mHz, while at the same time the $\vec{v}_+$ combination shows improvement of 12 $\%$ at low frequencies rising to 
over 50 $\%$ at 20 mHz.  
Finally, Fig.\ref{fig:sens-net} exhibits the sensitivities of the network observable over various source directions. 
Since the sensitivity of this observable is independent of $\phi_B$, we plot the curves for several values of 
$\theta_B = 0, 30, 60, 90$ degrees. Also since the network observable possesses reflection symmetry about the ecliptic plane 
$\theta_B = 90^{\circ}$, we do not need to plot the curves $\theta_B$ between $90^{\circ}$ and $180^{\circ}$. 
The important observation from this figure is that not much variation in sensitivity is seen as all source directions 
are scanned. Thus the network observable integrated over the year has essentially uniform sensitivity to all 
source directions over the frequency range $10^{-4} - 10^{-1} $ Hz. 

\section{Conclusion}

 We have shown how the SNR can be optimised for a GW source with known direction but with unknown polarisation. While 
obtaining the SNR the signal is averaged over the polarisations and then optimised. Because of this procedure we lose out 
to some extent on the SNR but on the otherhand it leads to robust results. The optimisation methods are algebraic 
in that one must solve an eigenvalue equation to determine the optimum SNRs. We separately deal with the low frequency 
case as it is of considerable astrophysical importance - a large fraction of GW sources are expected to be of this 
category. Secondly, it lends itself to simple analytical approximations which throw light on the results obtained.
Lastly, we deal with the general case covering the full band-width of LISA. We have compared the sensitivities obtained 
with our strategy to those obtained in the standard way. We find that the improvement in sensitivity of the network 
observable over the $X$ or $\X$ ranges from about 34\% to nearly 90\% over the bandwidth of LISA for a source lying the 
ecliptic plane. Finally we present a list of few binaries in our galaxy for which the optimal SNRs and the 
network SNRs have been computed. We also describe a method of extracting information about the inclination angle of the 
orbit of the binary if $SNR_{+, \times}$ can be measured. 

\section{acknowledgements}

The authors would like to thank the IFCPAR under which this work has been carried out.
A. Pai would like to thank the CNRS, Observatoire de la C\^ote d'Azur for
post-roug\'e position under which this work was carried out.

\end{document}